\documentclass[twocolumn,fontsize=9pt]{scrartcl}
\usepackage[a4paper, margin=0.5in]{geometry}
\usepackage{graphicx}
\usepackage{lmodern}
\usepackage[utf8]{inputenc}
\usepackage{amsmath}
\usepackage{amssymb}
\usepackage{fancyhdr}
\usepackage[usenames,dvipsnames,svgnames,table]{xcolor}
\usepackage{float}
\usepackage[tableposition=top,labelfont=bf,textfont=sf]{caption}
\usepackage{subcaption}
\usepackage{url}

\usepackage{upgreek}

\usepackage{siunitx} 
\usepackage{multirow} 
\usepackage{booktabs} 

\usepackage{lscape}
\usepackage{xstring}
\usepackage{tikz}
\usepackage[europeanresistors]{circuitikz}
\usepackage{pgfplots}
\pgfplotsset{width=0.9\textwidth, height=0.7\textwidth, compat=newest}
\usepackage{rotating}
\pgfkeys{/pgf/number format/relative round mode=fixed}
\usetikzlibrary{shapes.geometric,calc, arrows, shapes, decorations.markings, plotmarks}
\usepackage{adjustbox}
\usepackage{appendix}
\usepackage{xspace} 
\usepackage[explicit]{titlesec} 
\usepackage[onehalfspacing]{setspace}

\usepgfplotslibrary{groupplots,statistics} 
\pgfdeclarelayer{background}
\pgfsetlayers{background,main}

\makeatletter 
\g@addto@macro\UrlBreaks{
  \do\a\do\b\do\c\do\d\do\e\do\f\do\g\do\h\do\i\do\j
  \do\k\do\l\do\m\do\n\do\o\do\p\do\q\do\r\do\s\do\t
  \do\u\do\v\do\w\do\x\do\y\do\z\do\&\do\1\do\2\do\3
  \do\4\do\5\do\6\do\7\do\8\do\9\do\0}
\makeatother

\usepackage[minnames=1,maxnames=4,doi=false,isbn=false,style=nature,eprint=false,backend=biber,natbib=true,hyperref=true]{biblatex}

\definecolor{darkblue}{rgb}{0,0,.3}
\usepackage[bookmarksopen=true, bookmarksopenlevel=4, breaklinks=true, pdfstartview=Fit,colorlinks=true,
			linkcolor=darkblue, 
			citecolor=darkblue, 
			urlcolor=darkblue, 
			filecolor=darkblue]{hyperref}
\newcommand{\eqpoint}{\ensuremath{\ \ \ .}}
\newcommand{\m}{\,\si{\metre}}
\newcommand{\cm}{\,\si{\cm}}
\newcommand{\e}{\,\si{\elementarycharge}}

\newcommand{\kHz}{\,\si{\kHz}}
\newcommand{\MHz}{\,\si{\MHz}}
\newcommand{\mHz}{\,\si{\mHz}}

\newcommand{\Hz}{\,\ensuremath{\mathrm{Hz}}}

\newcommand{\s}{\,\ensuremath{\mathrm{s}}}
\newcommand{\K}{\,\si{\K}}

\newcommand{\ohm}{\,\si{\ohm}}

\newcommand{\mohm}{\,\si{\mohm}}

\newcommand{\pF}{\,\si{\pF}}

\newcommand{\mus}{\,\si{\micro\ensuremath{\mathrm{s}}}}

\newcommand{\tesla}{\,\si{\tesla}}
\newcommand{\kelvin}{\,\si{\kelvin}}

\newcommand{\minute}{\,\si{\minute}}
\newcommand{\hour}{\,\si{\hour}}

\newcommand{\trap}{\textsc{Pentatrap}\xspace}

\title{Detection of metastable electronic states by Penning-trap mass spectrometry}
   \small \author{\parbox{15cm}{R. X. Schüssler$^{1,*}$, H. Bekker$^{1, \dagger}$, M. Braß$^2$, H. Cakir$^1$, J. R. {Crespo López-Urrutia}$^1$, M. Door$^1$, P. Filianin$^1$, 
 Z. Harman$^1$, M. W. Haverkort$^2$, W. J.  Huang$^1$, P. Indelicato$^3$, C. H. Keitel$^1$, C. M. König$^1$, K. Kromer$^1$, M. Müller$^1$,
Yu. N. Novikov$^{4,5}$, A. Rischka$^1$, Ch. Schweiger$^1$, S. Sturm$^1$, S. Ulmer$^6$, S. Eliseev$^1$, K. Blaum$^1$}}
\date{%
    $^1$Max Planck Institute for Nuclear Physics, 69117 Heidelberg, Germany\\
$^2$Institute for Theoretical Physics, Heidelberg University, 69120 Heidelberg, Germany\\
$^3$Laboratoire Kastler Brossel, Sorbonne Université, CNRS, ENS-PSL Research University, Collège de France, Campus Pierre et Marie Curie, 4 place Jussieu, 75005 Paris, France\\
$^4$Petersburg Nuclear Physics Institute, 188300 Gatchina, Russia\\
$^5$St.Petersburg State University, 199304 St.Petersburg, Russia\\
$^6$RIKEN, Fundamental Symmetries Laboratory, Wako, Saitama 351-0198, Japan\\
Present addresses: $^{\dagger}$ Department of Physics, Columbia University, New York, NY, 10027-5255, USA
}
\usepackage{isotope}

\begin{document}
\definecolor{orangemovement}{RGB}{225,128,0} 
\maketitle

\textbf{State-of-the-art optical clocks \cite{Ludlow} achieve fractional precisions of $10^{-18}$ and below using ensembles of atoms in optical lattices \cite{katori2011optical,MartiPRL2018} or individual ions in radio-frequency traps \cite{brewer201927,nicholson2015systematic}. They are used as frequency standards and in searches for possible variations of fundamental constants \cite{safronova2018search}, dark matter detection \cite{Derevianko2014}, and  physics beyond the Standard Model \cite{lorentz,geodesy}. 
Promising candidates for novel clocks are highly charged ions (HCIs) \cite{KozlovRMP2018} and nuclear transitions \cite{Seiferle2019}, which are largely insensitive to external perturbations and reach wavelengths beyond the optical range \cite{nauta2017towards}, now becoming accessible to frequency combs \cite{cingoz2012direct}. However, insufficiently accurate atomic structure calculations still hinder the identification of suitable transitions in HCIs.\\
Here, we report on the discovery of a long-lived metastable electronic state in a HCI by measuring the mass difference of the ground and the excited state in Re, the first non-destructive, direct determination of an electronic excitation energy. This result agrees with our advanced calculations, and we confirmed them with an Os ion with the same electronic configuration. 
We used the high-precision Penning-trap mass spectrometer PENTATRAP, unique in its synchronous use of five individual traps for simultaneous mass measurements. 
The cyclotron frequency ratio $R$ of the ion in the ground state to the metastable state could be determined to a precision of $\delta R=1\cdot 10^{-11}$, unprecedented in the heavy atom regime.
With a lifetime of about 130 days, the potential soft x-ray frequency reference at $\nu=4.86\cdot 10^{16}\Hz$ has a linewidth of only $\Delta \nu\approx 5\cdot 10^{-8}\Hz$, and one of the highest electronic quality factor ($Q=\frac{\nu}{\Delta \nu}\approx 10^{24}$) ever seen in an experiment. 
Our low uncertainty enables searching for more HCI soft x-ray clock transitions \cite{CrespoHCI2016,nauta2017towards}, needed for promising precision studies of fundamental physics \cite{KozlovRMP2018} in a thus far unexplored frontier.
}

Modern clocks and frequency standards range from ensembles of neutral particles trapped in optical lattice clocks  \cite{ushijima2015cryogenic,nicholson2015systematic} to individual, singly charged ions confined in Paul traps \cite{Hunteman}.
With a fractional frequency accuracy of  $10^{-18}$ they allow performing excellent tests of fundamental symmetries, for example Lorentz invariance \cite{lorentz}, geodetic measurements \cite{geodesy}, and searching for new physics \cite{safronova2018search}.
The transitions used as frequency references typically have long lifetimes in the order of seconds, yielding sub-Hz linewidths; a recent example is the Yb$^+$ clock \cite{McGrew2018,lorentz}. Optical clocks use variations of the Ramsey method, driving such forbidden transitions through stimulated absorption and emission much faster than the natural transition rate, but maintaining, as needed for the interrogation schemes, very long coherence times due to the small spontaneous emission rate \cite{Ludlow}.
Possible promising species for a new generation of clocks are nuclear clocks \cite{Seiferle2019} or highly charged ions (HCIs) \cite{KozlovRMP2018}, since their compact size in comparison to atoms makes them less sensitive to external field fluctuations. 
Several transitions have been proposed to feature a high sensitivity to a variation of fundamental constants \cite{ong2014optical,Dzuba2015}.

While electronic binding energies in HCIs typically amount to several keV \cite{Gillaspy2001}, and inter-shell transitions usually appear in the x-ray region, there are also intra-shell fine and hyperfine transitions in the optical and UV range \cite{Klaft1994,Morgan1995,Crespo1996,schiller2007,Crespo2008}.
Furthermore, some HCIs feature level crossings \cite{berengut2012a,KozlovRMP2018} with associated optical transitions. 
Most of these transitions are E1-forbidden, many are of the M1-type, with lifetimes on the order of ms, but some are highly forbidden and can have extremely long lifetimes up to millions of years \cite{safronova2018search} as predicted theoretically. 
A few of them with lifetimes in the range from milliseconds to seconds have been found and investigated using storage rings \cite{SchippersPRL2007} and ion traps \cite{TraebertPRA2006,TraebertPRL2007}, but hitherto no method  allowed a direct observation of  the excitation energy of extremely long-lived metastable states.

It is difficult to theoretically predict such electronic energies in HCI. Calculations suffer from difficulties related to strong relativistic contributions and complex correlations of several active electrons in open shells. 
Experimental identification of the transitions on the other hand is challenging, as their long lifetime leads to sub-Hz linewidths and therefore requires very precise application of the Ritz-Rydberg method \cite{ritz1908new} to other accurately known transitions in order to establish the searching range to be explored with narrow-linewidth spectroscopy lasers.
Additionally, rough calculations with the flexible atomic code \cite{gu2008flexible} can be used to estimate the energy of transitions.

Here, we demonstrate for the first time how high-precision Penning-trap mass spectrometry directly identifies a suitable clock transition in HCIs by determining the mass difference of a $^{187}$Re$^{29+}$ ion in the ground state and a metastable state to a precision below 2\,eV or $\sim500$\,THz. 
A relative mass measurement of the two HCI states with an unprecedented precision of $1\cdot 10^{-11}$ is hereby achieved  by determining the cyclotron frequency $\nu_c=qB/(2\pi m)$ of the HCI with mass $m$ and charge $q$ in a strong magnetic field ($B\sim7\,\text{T}$) and a weak electrostatic harmonic potential of the Penning trap.
The cyclotron frequency is derived from measuring  the ion trap frequencies $\nu_+$, $\nu_z$ and $\nu_-$ (modified cyclotron (16\,MHz), axial (700\,kHz) and magnetron frequency (10\,kHz)) and applying the invariance theorem  $\nu_c=\sqrt{\nu_+^2+\nu_z^2+\nu_-^2}$ \cite{invariance}.\\

\begin{figure*}[t!]
\includegraphics[width=\textwidth]{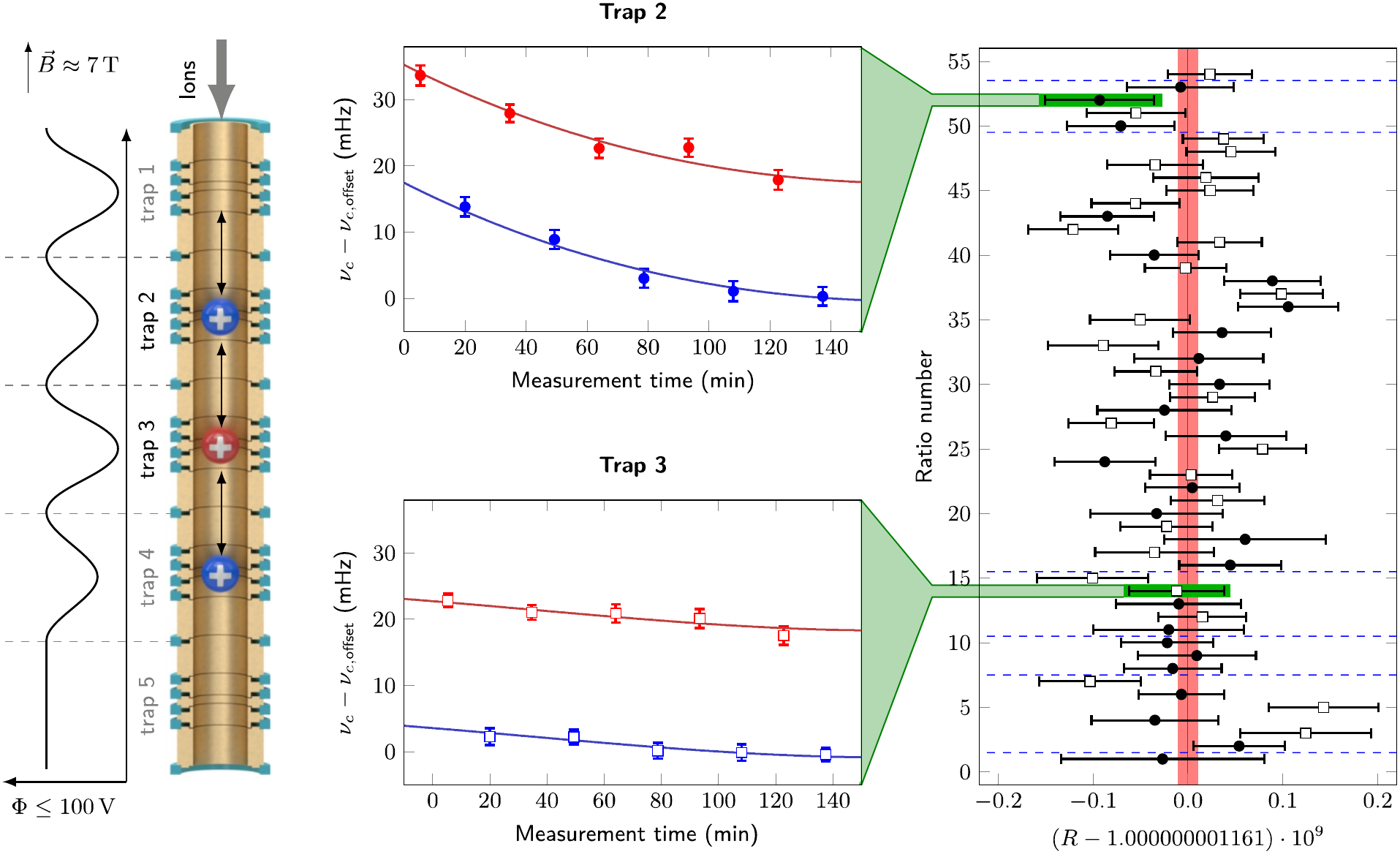}
\caption{Results of the Re measurements at \trap. Left: Three $^{187}\text{Re}^{29+}$ ions (ion 1: blue, ion 2: red and ion 3: blue) are loaded from the top into the trap stack. The  potentials $\Phi$ are nominally identical to the neighbouring traps for each of the measurement traps, i.e.\ the potential of trap 2 is also applied in trap 4. The magnetic field points along the direction of the trap axis.
Center: Shifting the ions one trap down or up after a cyclotron frequency determination results in measurements with the blue (ion 1) and red ion (ion 2) in trap 2 and red and second blue ion (ion 3) in trap 3. Here, the red ion is in the ground state, the other two are in the metastable electronic state. Right: All ratios determined over seven measurement campaigns (divided by the dashed lines) display the stability of the system. The results for trap 2 (filled) and trap 3 (empty) exhibit similar behaviour. The final averaged value is shown in red.}
\label{bild1}
\end{figure*}

This work was carried out with the novel high-precision Penning-trap mass spectrometer PENTATRAP at the Max-Planck-Institut f\"ur Kernphysik in Heidelberg.
HCIs produced in a commercial electron beam ion trap DreEBIT \cite{zschornack2009} using the MIVOC technique \cite{koivisto1998metal} are extracted in $1\mus$ long bunches with a kinetic energy of a few keV/q, selected according to their charge-to-mass ratio $q/m$ by means of a 90$^{\circ}$ dipole magnet, and sent into the Penning traps. 
Prior to the  trapping of the ions, their kinetic energy is reduced to a few eV/q by appropriately timed pulsed potentials applied to two cylindrical drift tubes. 
The 4\,K cold bore of the superconducting magnet houses five cylindrical Penning traps \cite{roux2012,repp2012}. 
Two of them (traps 2 and 3, see \autoref{bild1}) are used for measuring the trap frequencies of the ions of interest. 
Traps 1 and 4  serve to store ions, while trap 5 will allow monitoring fluctuations of the trap potentials and  the magnetic field in the future. 
In order to reduce temporal variations of the magnetic field in the traps, the temperature in the laboratory is kept constant on a level of 0.1\,K/day. 
Furthermore, the level of the liquid helium and the helium pressure inside the magnet cold bore are stabilised on a level of a fraction of a millimeter and 20\,$\upmu$bar, respectively. These measures reduce fractional changes of the magnetic field in the measurement traps to values below a few parts in $10^{10}$/h, resulting in a cyclotron-frequency drift of only a few\,mHz/h at $\nu_c=16\MHz$. 
Since  the magnet is actively shielded with a shielding factor of 50, the fractional fluctuation of the magnetic field in the trap due to the fluctuation of the stray magnetic field does not exceed 3 parts in 10$^{11}$.
Due to the extremely low pressure in the cryogenic trap, charge exchange with residual gas atoms is strongly suppressed, and trapped HCIs remain in the desired charge state over a couple of days.

The measurement procedure of the cyclotron-frequency ratio of two different ion species consists of preparatory and main phases. 
During the preparatory phase, three ions of two species are first loaded into the three innermost traps (ion 1 in trap 2 and ion 3 in trap 4 being in the same state, e.~g.\ the ground state, whereas ion 2 in trap 3 being in the other state). 
After that, their trap-motion amplitudes are reduced (the ions are ``cooled") by coupling the magnetron and cyclotron motions to the axial motion, which in turn is brought into thermal equilibrium with a tank circuit having a temperature around 6\,K. 
The resonance frequency  of the tank circuit is approximately equal to that of the axial motion. 
The axial, magnetron, and cyclotron amplitudes of cooled $^{187}$Re$^{29+}$ ions are approximately 10\,$\upmu$m, 2\,$\upmu$m, and 2\,$\upmu$m, respectively. 
Great care is taken to prepare  just a single ion of interest in each trap as the presence of another undesired ion in the trap would disturb the motion of the ion of interest and hence alter its motional frequencies.  
The main phase is devoted to the measurement of the ion trap frequencies in traps 2 and 3.       
The reduced cyclotron and axial frequencies are \textit{simultaneously}  measured by means of the Pulse-and-Phase (PnP) \cite{pnp} and dip \cite{doi:10.1063/1.360947} techniques, respectively. 
This minimises systematic shifts arising from changes in the external fields compared to sequential measurements of the trap frequencies.
The measurement cycle consists of a set of sub-measurements. First, the cyclotron motion is excited to an amplitude of approximately 10 $\upmu$m with a subsequent measurement of its phase (reference phase). 
It is followed by the re-cooling of all three motions. 
After that, the cyclotron motion is again excited to the same amplitude and is let to freely evolve its phase for 40\,s with its subsequent measurement (measurement phase). During the phase-evolution time of 40 s the axial frequency is measured with the dip method \cite{doi:10.1063/1.360947}. 
The measurement cycle is ended by re-cooling all ion motions. 
As the relative mass difference between the ground and metastable states is on the level of $10^{-9}$, it is sufficient to measure the smallest trap frequency, the magnetron frequency, once a day with a moderate precision for the determination of the cyclotron frequency ratio.
Ten measurement cycles constitute a measurement run.
A unique feature of PENTATRAP is that during each run, frequency measurements are performed $simultaneously$ in synchronized traps, for instance on ion 1 in trap 2 (upper blue ion in \autoref{bild1}) and ion 2 (red) in trap 3, respectively, for approximately 12\,min. 
After that, the ion species are swapped between traps 2 and 3 by moving the three ions one trap up. Then, a measurement run is carried out with ion 2 in trap 2 and ion 3 (lower blue ion) in trap 3. Lastly, the ion species are  swapped back by  moving the three ions one trap down again. 
After each change of the configuration of ions in the traps the ion motions are cooled for 20\,s. 
This measurement sequence is repeated until the measurement is stopped to empty the traps and load a new set of ions.
In this way, one alternately determines the cyclotron frequencies of ions in ground and metastable states in each measurement trap.

In order to obtain the cyclotron-frequency ratio of the ions in ground and metastable states, the polynomial method described, for instance, in \cite{PhysRevLett.115.062501,PhysRevC.100.015502,PhysRevA.72.022510} is applied in the following way. 
We treat the two measurement traps independently.
The temporal variation of the magnetic field and hence, the cyclotron frequency is approximated with a low-order polynomial, as shown  for both measurement traps in the plots in the middle of \autoref{bild1}.
This assumption is based on our experience with the magnetic-field evolution over a few-hours scale. 
The full set of measured cyclotron frequencies is divided in a set of groups.  
The group length must be as short as possible such that it can be described by a low-order polynomial, but must contain substantially more data points than the polynomial degrees of freedom. 
Thus, each group is chosen to be approximately 2-hour long and to contain a total of 10 cyclotron-frequency points (five points correspond to the ion in the ground state and the other five points correspond to the ion in the metastable state), see middle of \autoref{bild1}. 
We use third-grade polynomials since this is the lowest order that has at least one inflection point.
Data points from the ions in both the metastable and the ground state are fitted with the same polynomial with global fitting parameters.
The polynomial for the data points from the ion in the ground state is scaled with an additional fitting parameter $R$, which corresponds to the frequency ratio.
This yields three quantities: (1) the frequency ratio $R$, (2) its uncertainty $\delta R$ and (3) the reduced $\chi^2$. 
If the latter is greater than one, we scale the uncertainty of the ratio by $\chi^2$. 
The final ratio $R$ is obtained as a weighted mean  of the frequency ratios from all groups and measurement traps. 
Its statistical uncertainty $\delta R$  is the larger one of the internal and external errors \cite{Birge}, see right side of \autoref{bild1}. 
A variation of the polynomial order (second, third and forth ) results in an increase of the statistical uncertainty of the final ratio by approximately 10$\%$. 
The data were  blindly analysed independently by two people, each  using separately developed software. 
Both analyses yield similar external errors of $8 \cdot 10^{-12}$ and internal errors of $7 \cdot 10^{-12}$ resulting in a Birge ratio of 1.14. 
The two weighted means differ by $4 \cdot 10^{-12}$. 
Thus, the combined final ratio is  
\begin{equation}
    R-1=1.161(10)_{\textrm{\tiny stat}}\cdot 10^{-9} \eqpoint
\end{equation}
The uniqueness of the considered $^{187}$Re$^{29+}$ ions in two electron-configuration states is that the fractional difference in their masses is just 10$^{-9}$. 
Uncertainties due to systematic effects such as image charge shift, relativistic shift, and higher order magnetic and electric field (see for example \cite{Smorra2017}) variations can be neglected. For these to be on the level of $10\%$ of the statistical uncertainty ($10^{-12}$), the cyclotron frequencies would have to be shifted by at least 10\,kHz, which is excluded for the present known performance of PENTATRAP.
Therefore, the systematic uncertainties  in the final ratio are well below the statistical uncertainty and hence can be neglected.

The ground and metastable states in $^{187}$Re$^{29+}$ belong to the $\left[\textrm{Kr}\right]4\textrm{d}^{10}$~$^1S_0$ and $\left[\textrm{Kr}\right]4\textrm{d}^{9}4\textrm{f}^{1}$~$^3H_5$ electron configurations, respectively. 
Since the metastable one can only decay to the ground state, and this by a highly forbidden electric triacontadipole (E5) transition, its estimated lifetime is $\sim130\,\textrm{d}$.
Due to repeated and fast processes of electron-impact excitation followed by radiative decay inside the EBIT, low-lying metastable states and the ground state are more or less equally populated. 
In the present case, the metastable ion fraction detected in PENTATRAP is approximately 50\%.
For the determination of the energy of the metastable state, 7 measurement campaigns with three ions each were performed over the course of 12 days, during which not a single decay was detected.

From the final ratio, the energy of the metastable state with respect to the ground state can be calculated using $\Delta E_{\textrm{Re}}=m\left(^{187}\textrm{Re}^{29+}\right)(R-1)c^2$.
The uncertainty of the ion mass  enters into the total uncertainty  reduced by a factor  $R-1$ and has to be known to only $\delta m/m\approx 10^{-4}$ to become negligible. 
Since the mass of the removed 29 electrons and their binding energies is well below that limit, we can simply take the neutral mass of $^{187}$Re for the calculation, giving us $\Delta E_{\textrm{Re}}=202.2(17)$\,eV.

\begin{figure}[t!]
\includegraphics[width=0.95\columnwidth]{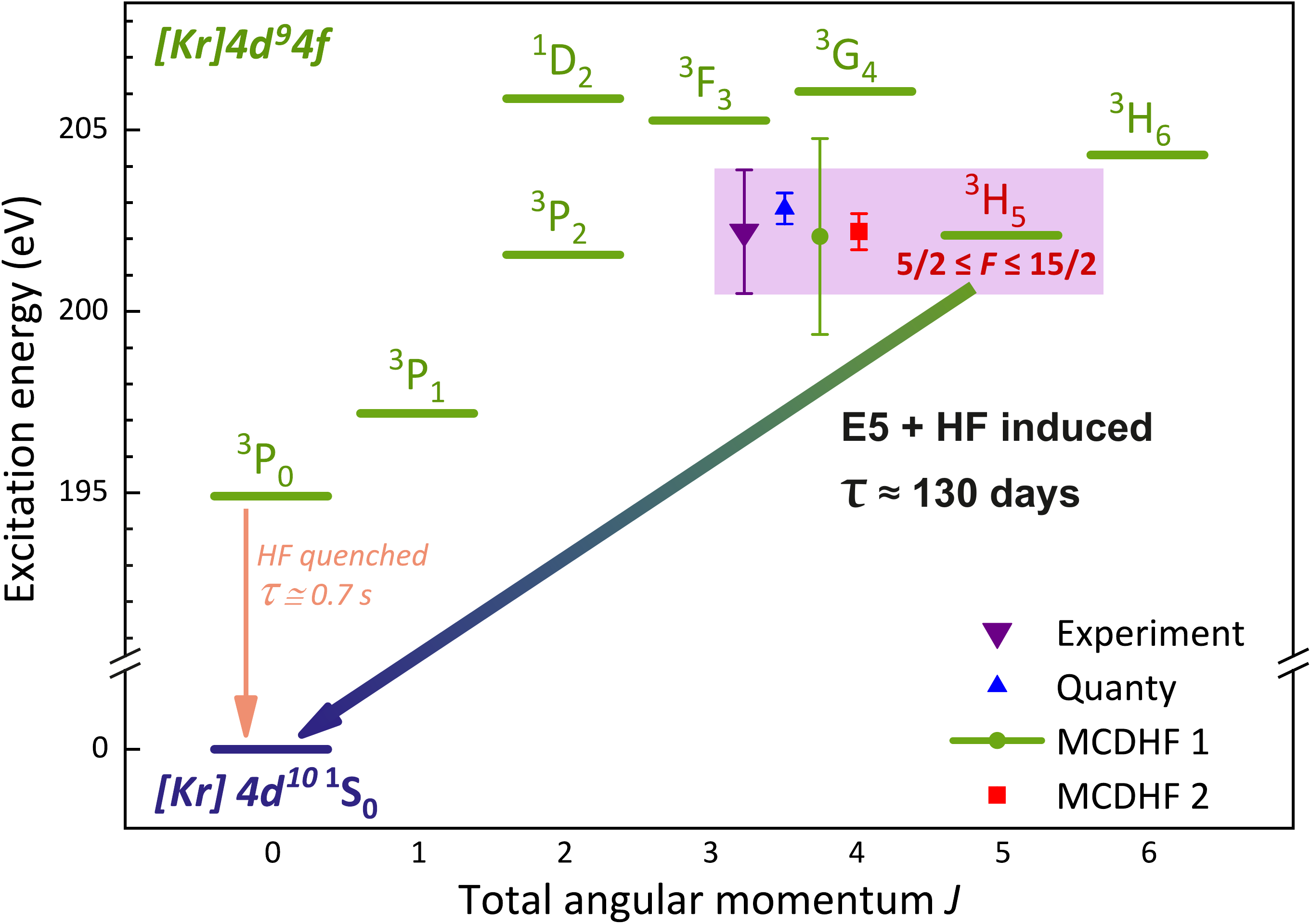}
\caption{The $4d^{10}$ ground state and relevant $4d^{9}4f$ excited electronic states of the $^{187}$Re${}^{29+}$ ion. Comparison of the experimental result  and theoretical values obtained using multi-configuration Dirac-Hartree Fock approaches in two different implementations (MCDHF 1 and 2) and by means of a configuration-interaction (Quanty) calculation is shown in the shaded region, which shows the experimental result.}
\label{levelscheme}
\end{figure}

The high charge state and the complexity of the open $4d$ and $4f$ subshells calls for state-of-the-art theoretical methods to match the experimental accuracy. We apply three different fully relativistic approaches, namely, the multiconfiguration Dirac-Hartree-Fock (MCDHF) method~\cite{Grant1970,Desclaux1971} in two different implementations, and the configuration interaction   method employing Kohn-Sham orbitals, as implemented by the many-body script language Quanty~\cite{Haverkort2016hz}.
The calculations also help to identify the observed metastable state. The first such state could be the $4d^{9}4f$~$^3P_0$ level with an energy of $\approx 195\,\textrm{eV}$ (see \autoref{levelscheme}). This level may only decay to the ground state by a highly forbidden two-photon transition. However, $^{187}$Re has a nuclear spin $I=5/2$ and thus a magnetic moment, which couples this $^3P_0$ level to the nearby $^3P_1$ level, which can decay to the ground state by an electric dipole E1 transition. The  $^3P_0$ state thus acquires part of the probability of this allowed transition, a phenomenon called \emph{hyperfine quenching} \cite{ipm1989,gmm1974,ibbc1992,bbcd1993}. 
Following Ref.\ \cite{ipm1989} and by employing the MCDFGME code~\cite{Indelicato2005} we estimate a lifetime of \SI{0.7}{\second}, which is too short to be observed. The observed state must then be the next metastable level, $4d^{9}4f$~$^3H_5$, which, neglecting quenching, can decay to the ground state only by an $E5$ transition with a lifetime of $\approx 239\,\textrm{d}$. Decay to the levels just below, i.e., to the $^3P_0$, $^3P_1$ and $^3P_2$ states, have very low probabilities because of their high multipolarities (M5, E4 and M3, respectively) and small photon energies. The $^3H_5$ state is also subject to hyperfine quenching, e.g. its lifetime with a coupled atomic and nuclear angular momentum $F=7/2$ reduces to $\approx 124\,\textrm{d}$. This  corresponds to a width-to-transition energy ratio of $3\cdot 10^{-25}$, allowing for very high-precision measurements. The states with $F=9/2$ and $F=11/2$ live \SI{128}{\day} and the other ones, which are coupled to more distant fine-structure levels, have even longer lifetimes.

In the Quanty calculations of transition energies, the relativistic Dirac Hamiltonian includes the Coulomb and Breit interactions between all electrons. The Fock space is spanned by multi-Slater-determinant
states constructed from relativistic Kohn-Sham single-electron states, obtained from the density functional code FPLO \cite{Koepernik1999uw}, using the local spin density functional of Ref.~\cite{Perdew1992}. Within this calculation we explicitly include shells with principal quantum numbers $n<n_{max}=7$. In order
to extrapolate to $n_{max}=\infty$ we compare results for calculations with $n<7$, $n<6$, and $n<5$. Allowing quantum fluctuations within the $n=4$ shells, i.e.,~ including configurations with holes in the $4s-4d$ shells and additional electrons in the $4f$ shell, we calculate the energies of the $4d^{9}4f$ excited states in a first approximation. Allowing for a maximum of three electrons in the $4f$ shell suffices to calculate the excitation energy with the required accuracy. Next, we extend the configuration space to allow electrons to scatter between the $n=$ 4, 5 and 6 shells, whereby at most three holes are excited.
By successively adding further configurations, we can extrapolate the excitation energy of the $^3H_5$ metastable state to be between 202.41 and 203.26 eV.

In the MCDHF calculations, we do not only optimize the mixing coefficients of the configurations but also the radial single-electron wave functions by self-consistently solving the corresponding radial Dirac equations.
QED corrections to the excitation energy were estimated by the model potential method~\cite{Shabaev2015} and by computing the self-energy shift of the $4d$ and $4f$ valence electrons employing effective potentials, which account for the screening by the core electrons.
One of the MCDHF calculations (MCDHF1) employs the MCDFGME~\cite{Indelicato2005}, including full relaxation of all spectroscopic orbitals, and single and double excitations from all $n=4$ orbitals to the free single-electron states up to $5d$, with a result of 202.1(27)\,eV. 
In the other computation (MCDHF2), employing {GRASP2018}~\cite{GRASP2018}, we generate the set of configurations with single and double excitations from the $n=4$ states up to $9h$, with the virtual orbitals optimized layer by layer. Subsequently, the effect of triple electron exchanges is accounted for in a configuration interaction approach. The convergence of the results while systematically expanding the configuration set is monitored, allowing to estimate the uncertainty of the calculations, and arriving to an extrapolated value of \SI{202.2+-0.5}{\electronvolt}.




\autoref{levelscheme} shows the comparison of the three theoretical values with the experimental one.
As a crosscheck we also performed, with less statistics, frequency-ratio measurements of the same metastable state in $^{187}\textrm{Os}^{30+}$, which is isoelectronic to $^{187}\textrm{Re}^{29+}$. 
We obtain $\Delta E_{\textrm{Os}}=207(3)$\,eV. 
As the electronic binding energy roughly scales with $Z^2$, this higher energy is explained by the additional proton in Os ($Z=76$) and the experimental result is again in agreement with the theoretical predictions, which are not explicitly given.

In summary, we have demonstrated a new way to determine the excitation energies of extremely long-lived metastable states in HCIs, which yields realistic measurement times if the metastable fraction in the atomic or ionic sample is at least 10\% and the lifetime longer than a few hours.
In principle, with the precision achieved here for Re ions it would also be possible to detect nuclear transitions with energies as low as 5\,eV, if the abundance of the isomeric nuclear state in the ions is comparable to that of the ground state.

\textbf{Data availability}

The datasets analysed for this study will be made available on reasonable request.

\textbf{Code  availability}

The  analysis codes will be made available on reasonable request.

\printbibliography

\textbf{Acknowledgements}
This article comprises
parts of the PhD thesis work of R.\,X.\,S. and H.\,C.
This work is part of and supported by the German Research Foundation (DFG) Collaborative Research Centre “SFB 1225 (ISOQUANT)'' and by the DFG Research UNIT FOR 2202.
P.\,I. acknowledges partial support from NIST. Laboratoire
Kastler Brossel (LKB) is “Unit\'{e} Mixte de Recherche de
Sorbonne Universit\'{e}, de ENS-PSL Research University, du Coll\'{e}ge de France et du CNRS n$\circ$ 8552”. 
P.\,I., Yu.\,N. and K.\,B. are members of the Allianz Program of the Helmholtz Association, contract n$\circ$ EMMI HA-216 “Extremes of Density and Temperature: Cosmic Matter in the Laboratory”. 
P.\,I. wishes to thank Jean-Paul Desclaux for his help improving the MCDFGME code. 
This project has received funding from the European Research Council (ERC) under the European Union’s Horizon 2020 research and innovation programme
under grant agreement No.\ 832848 - FunI. Furthermore,
we acknowledge funding and support by the International Max Planck Research School for Precision Tests of Fundamental Symmetries (IMPRS-PTFS) and by the Max Planck, RIKEN, PTB Center for Time, Constants and Fundamental Symmetries.

\textbf{Author contributions} 
The experiment was performed by R.X.S., M.D., A.R. and S.E.
The data was analysed by R.X.S. and S.E.
Theoretical calculations were performed by H.B., M.B., H.C., Z.H., M.H. and P.I.
The manuscript was written by R.X.S., M.B., Z.H., M.H., J.R.C.L.U. and S.E. and edited by S.U. and K.B.
All authors discussed and approved the data as well as the manuscript.
\end{document}